\documentclass{article}

\usepackage{amssymb}
\usepackage[english]{babel}
\usepackage{bm}
\usepackage{booktabs}
\usepackage{calc}
\usepackage[font=footnotesize,skip=5pt]{caption}
\usepackage{chemformula}
\usepackage{color}
\usepackage{commath}
\usepackage[shortlabels]{enumitem}
\usepackage[T1]{fontenc}
\usepackage{float}
\usepackage{gensymb}
\usepackage{geometry}
\usepackage{graphicx}
\usepackage{hanging}
\usepackage{hyperref}
\usepackage{lipsum}
\usepackage{mathrsfs}
\usepackage{mathtools}
\usepackage{multicol}
\usepackage{multirow}
\usepackage[square,numbers,compress]{natbib}
\usepackage{soul}
\usepackage{textcomp}
\usepackage[small]{titlesec}
\usepackage{url}
\usepackage{wrapfig}
\usepackage{setspace}

\usepackage[mathlines]{lineno} 
\usepackage{amsmath}           
\usepackage{etoolbox}          

\newcommand*\linenomathpatch[1]{%
  \cspreto{#1}{\linenomath}%
  \cspreto{#1*}{\linenomath}%
  \csappto{end#1}{\endlinenomath}%
  \csappto{end#1*}{\endlinenomath}%
}
\newcommand*\linenomathpatchAMS[1]{%
  \cspreto{#1}{\linenomathAMS}%
  \cspreto{#1*}{\linenomathAMS}%
  \csappto{end#1}{\endlinenomath}%
  \csappto{end#1*}{\endlinenomath}%
}

\expandafter\ifx\linenomath\linenomathWithnumbers
  \let\linenomathAMS\linenomathWithnumbers
  \patchcmd\linenomathAMS{\advance\postdisplaypenalty\linenopenalty}{}{}{}
\else
  \let\linenomathAMS\linenomathNonumbers
\fi

\linenomathpatch{equation}
\linenomathpatchAMS{gather}
\linenomathpatchAMS{multline}
\linenomathpatchAMS{align}
\linenomathpatchAMS{alignat}
\linenomathpatchAMS{flalign}

\makeatletter
\patchcmd{\mmeasure@}{\measuring@true}{
  \measuring@true
  \ifnum-\linenopenaltypar>\interdisplaylinepenalty
    \advance\interdisplaylinepenalty-\linenopenalty
  \fi
  }{}{}
\makeatother

\newcommand{\D}{\mathrm{d}}

\definecolor{newred}{RGB}{180,20,5}

\definecolor{newlight green}{RGB}{1,129,30}

\definecolor{newblue}{RGB}{40,100,250}

\begin{document}
\twocolumn[
{\centering
\vskip 1cm
{\large \bf{Probing growth precursor diffusion lengths by inter-surface diffusion}}
%
\vskip 0.5cm
Stoffel D.\ Janssens$^{1,\text{\textasteriskcentered}}$, Francisco S.\ Forte Neto$^2$, David V\'azquez-Cort\'es$^1$, Fernando P.\ Duda$^2$, and Eliot Fried$^{1,\text{\textasteriskcentered}}$
\vskip 0.5cm
$^1$Mechanics and Materials Unit (MMU), Okinawa Institute of Science and Technology Graduate University (OIST), 1919-1 Tancha, Onna-son, Kunigami-gun, Okinawa, Japan 904-0495\\
$^2$Programa de Engenharia Mec\^anica, COPPE, Universidade Federal do Rio de Janeiro, Cidade Universit\'aria, Rio de Janeiro/RJ, Brazil 21941-972

\vskip 0.5cm
$^{\text{\textasteriskcentered}}$Corresponding authors: Stoffel D.\ Janssens (stoffel.janssens@oist.jp), Eliot Fried (eliot.fried@oist.jp)
}

\vskip 0.5cm
Understanding and optimizing thin-film synthesis requires measuring the diffusion length $d_\alpha$ of adsorbed growth precursors. Despite technological advances, in-situ measurements of $d_\alpha$ are often unachievable due to harsh deposition conditions, such as high temperatures or reactive environments. In this paper, we propose a fitting approach to determine $d_\alpha$ from experimental data by leveraging inter-surface diffusion between a substrate and a strip obtained by, for example, processing a film. The substrate serves as a source or sink of precursors, influencing the growth dynamics and shaping the profile of the strip. By fitting simulated profiles to given profiles, we demonstrate that $d_\alpha$ can be determined. To achieve this, we develop a theoretical growth model, a simulation strategy, and a fitting procedure. The growth model incorporates inter-surface diffusion, adsorption, and desorption of growth precursors, with growth being proportional to the concentration of adsorbed precursors. In our simulations, a chain of nodes represents a profile, and growth is captured by the displacement of those nodes, while keeping the node density approximately constant. For strips significantly wider than $d_\alpha$, a scaled precursor concentration and $d_\alpha$ are the fitting parameters that are determined by minimizing a suitably defined measure of the distance between simulated and given profiles. We evaluate the robustness of our procedure by analyzing the effect of profile resolution and noise on the fitted parameters. Our approach can offer valuable insights into thin-film growth processes, such as those occurring during plasma-enhanced chemical vapor deposition.

\vskip 1cm
]

\section{Introduction}
Probing the atomic or molecular processes at the surface of a growing film or particle is challenging to achieve experimentally \cite{Allen1996}. Techniques like time-resolved environmental transmission electron microscopy (TEM) and in situ scanning probe microscopy provide valuable insights into these processes \cite{Tao2011,Alcorn2023,Wang2024}. However, the conditions required for material growth often differ from those that can be accommodated by in-situ characterization techniques. Therefore, novel technologies and approaches that enable the investigation of surface processes in harsh environments, such as those found in plasma-enhanced chemical vapor deposition \cite{Yi2021,Giussani2022,Teraji2024,Berger2024}, are highly desired to understand and improve thin-film synthesis. 

To theoretically understand thin-film synthesis, simulations based on density functional theory \cite{Guillaume2024,Li2024}, molecular dynamics \cite{Chen2024,Fernandez2024}, and the Monte-Carlo method \cite{Jensen1999,Rodgers2015,Luis2020,Valentin2022} are employed at the nanometer scale. The results from these simulations can serve as parameters, such as reaction rates, for continuum models \cite{Cukier1983,Wolf1990,Jabbour2003,Hu2024}, which in turn can predict growth at larger scales. Growth model parameters can also be estimated through a fitting procedure, in which discrepancies between simulated and experimental results are iteratively minimized. \citet{Bouchet2002} applied this approach to extract composition-dependent interdiffusion coefficients from concentration profiles. Additionally, the growth of systems with multiple objects, such as grains coalescing during growth, can be simulated using network theory \cite{Janssens2022}, the mean-field approach \cite{Tomellini2023}, or the level set method \cite{Smereka2005}.

In this paper, we introduce a fitting approach to determine the diffusion lengths of adsorbed growth precursors during thin film synthesis. We exploit the inter-surface diffusion of these precursors between a substrate and a strip obtained, for example, by film processing. The substrate can act as either a source or a sink for the precursors, influencing the precursor concentration near the strip--substrate--fluid triple line. Consequently, the growth of the strip becomes inhomogeneous, which is reflected in its profile. In \secref{sec:theoretical_model}, we develop a theoretical growth model and show that for relatively wide strips, the evolution of the profile of the strip is determined by (i) the diffusion length of the precursors on the strip and (ii) the precursor concentration at the triple line, relative to that at the center. Towards demonstrating that these parameters can be obtained by fitting simulated profiles to given profiles, we develop a simulation strategy in \secref{sec:sim_strat} and a fitting procedure in \secref{sec:fit}. In \secref{sec:demo}, we demonstrate our approach and provide a strategy for dealing with noisy given profiles.

Inspiration for this paper comes from experimental observations of inter-surface diffusion. Recently, we found evidence for the inter-surface diffusion of growth precursors from a silicon substrate to a diamond particle \cite{Cortes2024}. Our plasma-enhanced chemical vapor deposition experiments suggest that the silicon substrate is a source of growth precursors. Similar evidence can be inferred from the works of \citet{Park1994}, \citet{Anupam2021}, and \citet{Sartori2019}. Next to impacting the morphology of diamond films significantly, changes in the precursor concentration can substantially impact doping and, subsequently, electronic properties \cite{Janssens2011,Rouzbahani2021}. Using a combination of reflection high-energy electron diffraction, scanning electron microscopy (SEM), and molecular beam epitaxy, \citet{Nishinaga2005} and co-workers demonstrated that inter-surface diffusion between crystal surfaces with different crystallographic orientations also plays a significant role in crystal growth. Similar conclusions were reached by \citet{Xia2013} while studying noble metal crystal profiles with TEM. From \citet{Yamaguchi1992}, we infer that inter-surface diffusion also occurs between thin-film masks and crystals during metal-organic chemical vapor deposition.

\section{Theoretical growth model}
\label{sec:theoretical_model}
\begin{figure}
    \centering
    \includegraphics{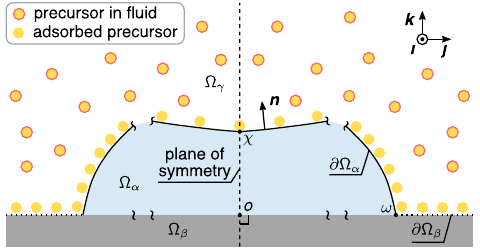}
    \caption{{\bf Theoretical growth model.} Cross-section schematic of strip $\Omega_\alpha$ that is growing on flat solid substrate $\Omega_\beta$ in fluid $\Omega_\gamma$. Surfaces $\partial\Omega_\alpha$ and $\partial\Omega_\beta$ of $\Omega_\alpha$ and $\Omega_\beta$, respectively, adsorb precursors that diffuse along and between $\partial\Omega_\alpha$ and $\partial\Omega_\beta$. At a point on $\partial\Omega_\alpha$, growth occurs in the direction of the surface unit normal vector $\bm{n}$ defined at that point and is proportional to areal particle concentration $\sigma$ of adsorbed growth precursors. The dashed line represents the plane of symmetry of $\Omega_\alpha$ and $\Omega_\beta$. Origin $o$ is that of a rectangular Cartesian coordinate system with axes $x$, $y$, and $z$ corresponding to orthonormal basis vectors $\bm{\imath}$, $\bm{\jmath}$, and $\bm{k}$. The triple line $\omega$ marks the intersection of $\partial\Omega_\alpha$, $\partial\Omega_\beta$, and $\Omega_\gamma$, and line $\chi$ lays at the intersection of $\partial\Omega_\alpha$ and the $xz$ plane.}
    \label{fig:model}
\end{figure}

We model the growth of strip $\Omega_\alpha$ that lies on flat solid substrate $\Omega_\beta$, as illustrated schematically in Fig.~\ref{fig:model}. Fluid $\Omega_\gamma$ provides growth precursors and is in contact with surfaces $\partial\Omega_\alpha$ and $\partial\Omega_\beta$ of the strip and the substrate, respectively. Adsorbed precursors can diffuse along and between $\partial \Omega_\alpha$ and $\partial \Omega_\beta$, and at a point on $\partial \Omega_\alpha$, growth proceeds in the direction of surface unit normal vector $\bm{n}$ defined there. The strip and substrate share a plane of symmetry that lies on the $xz$ plane of a rectangular Cartesian coordinate system with origin $o$ and coordinate axes $x$, $y$, and $z$ that correspond to the orthonormal basis vectors $\bm{\imath}$, $\bm{\jmath}$, and $\bm{k}$, respectively. Origin $o$ is located at the strip--substrate interface, and cross-sections of the strip perpendicular to $\bm{\imath}$ are identical. Due to symmetry, we only generate results on the side of the positive $x$ axis, where $\omega$ represents the strip--substrate--fluid triple line. Line $\chi$ marks the intersection of $\partial\Omega_\alpha$ and the $xz$ plane.

The etching of $\Omega_\alpha$ and $\Omega_\beta$, nucleation on $\Omega_\beta$, and effects arising from surface curvature are neglected. We assume that mass transport in $\Omega_\gamma$ is much faster than surface diffusion and that surface diffusion is much faster than growth (quasi-steady state approximation). The precursor concentration in $\Omega_\gamma$ is taken to be homogenous, and we assume Fickian diffusion and isothermal conditions. Reactions are captured by a modified Langmuir model that accounts for adsorption, desorption, and growth. 

In the Langmuir model, precursor P$_\gamma$ from $\Omega_\gamma$ occupies vacancy V$_\alpha$ on $\partial\Omega_\alpha$ to form adsorbed precursor P and occurs at rate $J^+_1$. Conversely, desorption occurs at rate $J^-_1$, producing P$_\gamma$ and V$_\alpha$ from P. Growth is modeled by the incorporation of P into $\Omega_\alpha$ and the formation of V$_\alpha$, and occurs at rate $J^+_2$. Reactions are represented by
\begin{equation}
\ch{$\text{P}_\gamma$ + $\text{V}_\alpha$ <=> P -> $\Omega_\alpha$ + $\text{V}_\alpha$} \quad \text{on} \quad \partial\Omega_\alpha.
\label{eq:chem_D}\\
\end{equation}
Similarly, adsorption and desorption, with rates $J^+_3$ and $J^-_3$, respectively, are represented by
\begin{equation}
\ch{$\text{P}_\gamma$ + $\text{V}_\beta$ <=> P} \quad \text{on} \quad \partial\Omega_\beta,
\label{eq:chem_S}
\end{equation}
where V$_\beta$ denotes a vacancy on $\partial \Omega_\beta$. Relations for $J_1^+$, $J_1^-$, $J_2^+$, $J_3^+$, and $J_3^-$ can be found in Appendix A.

On assuming that $J_2^+$ follows a first-order reaction rate, the growth rate $\mathcal{G}$ at any point on $\partial \Omega_\alpha$ is proportional to $\sigma$. For a sufficiently wide strip, boundary effects can be neglected at its center. At this location, $\mathcal{G} = \mathcal{G}_\alpha$ and $\sigma = \sigma_\alpha$. This allows us to express $\mathcal{G}$ as
\begin{equation}
\mathcal{G} = \frac{\sigma}{\sigma_\alpha} \mathcal{G}_\alpha,
\label{eq:growth_rate}
\end{equation}
where $\mathcal{G}_\alpha$ can be obtained from a reference growth on a sufficiently wide strip or film \cite{Cortes2023}. To determine $\sigma/\sigma_\alpha$, we balance masses \cite{Kolmogorov1937,Crank1975,Duda2023} and obtain
\begin{align}
D_\alpha \frac{\partial^2 \sigma}{\partial r^2} + J_1^+ - J_1^- - J_2^+ &= 0
&\text{on} \quad &\partial \Omega_\alpha \quad &\text{and}
\label{eq:KPP_D}\\
D_\beta \frac{\partial^2 \sigma}{\partial r^2} + J_3^+ - J_3^- &= 0
&\text{on} \quad &\partial\Omega_\beta,
\label{eq:KPP_S}
\end{align}
where $D_\alpha$ and $D_\beta$ are the diffusion coefficients of the adsorbed precursors and $r$ denotes arc length measured along $\partial \Omega_\alpha$ and $\partial \Omega_\beta$ from $\chi$ towards $\omega$ and beyond. Relations \eqref{eq:KPP_D} and \eqref{eq:KPP_S} are also referred to as reaction-diffusion equations \cite{Papp2024}.

General solutions of \eqref{eq:KPP_D} and \eqref{eq:KPP_S} are obtained following the method of combination of variables \cite{Boltzmann1894}. Subsequently, particular solutions are determined by incorporating the coupling condition that accounts for inter-surface diffusion, along with the relevant symmetry and boundary conditions. This procedure is delineated in Appendix A. On $\partial \Omega_\alpha$, we find
\begin{equation}
\frac{\sigma}{\sigma_\alpha} = 1 + \frac{2\left(\dfrac{\sigma_\omega}{\sigma_\alpha} - 1\right)\exp{\left(\dfrac{r_\omega}{d_\alpha}\right)}}
{\exp\left(\dfrac{2r_\omega}{d_\alpha}\right)+1}
\cosh{\left(\frac{r}{d_\alpha}\right)},
\label{eq:res_sol_r_D}
\end{equation}
in which $\sigma_\omega$ and $r_\omega$ are the values of $\sigma$ and $r$ at $\omega$, respectively, and $d_\alpha$ is the diffusion length of P. From the coupling condition, we find that
\begin{equation}
\frac{\sigma_\omega}{\sigma_\alpha} = \frac{\dfrac{\sigma_\beta}{\sigma_\alpha}+\dfrac{d_\beta}{d_\alpha}\dfrac{D_\alpha}{D_\beta}\tanh{\left(\dfrac{r_\omega}{d_\alpha}\right)}} 
{C+\dfrac{d_\beta}{d_\alpha}\dfrac{D_\alpha}{D_\beta}\tanh{\left(\dfrac{r_\omega}{d_\alpha}\right)}},
\label{eq:res_bcon}
\end{equation}
where $\sigma_\beta$, $d_\beta$, and $C$ are $\sigma$ on $\partial \Omega_\beta$ far away from $\omega$, the diffusion length of growth precursors in $\partial \Omega_\beta$, and a term including chemical potentials.

The parameters in \eqref{eq:res_sol_r_D} can be grouped in a minimum of four parameter groups, namely $\sigma_\beta/\sigma_\alpha$, $d_\beta D_\alpha / d_\alpha D_\beta$, $C$, and $d_\alpha$. With $r_\omega/d_\alpha$ sufficiently large, the parameters in \eqref{eq:res_sol_r_D} can be grouped in a minimum of two parameter groups, namely, $d_\alpha$ and $\acute{\sigma}_\omega/\sigma_\alpha$, where $\acute{\sigma}_\omega$ is the value of $\sigma_\omega$ when $\tanh{(r_\omega/d_\alpha)} = 1$.

\section{Simulation strategy}
\label{sec:sim_strat}
By multiplying $\mathcal{G}$ with time step $\D t$, we obtain the growth $\D g$ along $\bm{n}$ of which the discretized version has to be computed for our simulations. Similarly, $\mathcal{G}_\alpha \D t = \D g_\alpha$. With $\eqref{eq:growth_rate}$, we then find that
\begin{equation}
\D g = \frac{\sigma}{\sigma_\alpha} \D g_\alpha.
\label{eq:gr_rts}
\end{equation}
Relation \eqref{eq:gr_rts} allows us to use $g_\alpha$ as a growth parameter instead of $t$. For convenience, we take $g_\alpha = 0$ pre-growth and $g_\alpha = g_e$ post-growth, with $g_e$ denoting the thickness added by growth to the center of a reference film or strip taken sufficiently wide so that boundary effects can be negligible.

If for a certain strip with $r_\omega / d_\alpha = 5$ pre-growth and $\sigma_\omega / \sigma_\alpha = 2$, $\D g(\chi) \approx 1.01 \D g_\alpha$. Under these conditions, it can be reasonable to assume that the total thickness added to that strip at $\chi$ by growth is $g_e$. Then, $g_e$ can be obtained without a reference growth. Note that the center of a strip can be used as a reference point for comparing the profiles of a strip before and after growth.

\begin{figure}
    \centering
    \includegraphics{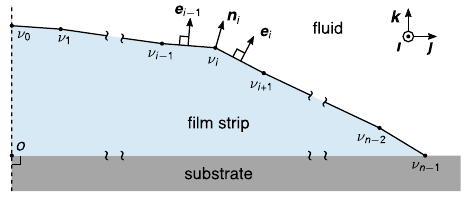}
    \caption{{\bf Simulation strategy.} Profile $\partial \Omega_\alpha$ represented by a chain of $n$ nodes, namely $v_i$, with $i = 0, \dots, n-1$. Except for $\nu_0$ and $\nu_{n-1}$, $\nu_i$ has an outward surface unit normal $\bm{n}_i$ that is a function of edge normals $\bm{e}_{i-1}$ and $\bm{e}_{i}$. During a growth step, $\nu_i$ moves in the direction of $\bm{n}_i$.}
   \label{fig:sim_strat}
\end{figure}

To simulate growth, we model the profile of $\partial\Omega_\alpha$ as a chain of $n$ nodes, as illustrated in Fig.~\ref{fig:sim_strat}. A similar representation can be used for moving boundary problems involving flows \cite{Sakakibara2021}. In our work, a node is defined by its $y$ and $z$ coordinates and is denoted as $\nu_i$, with $i = 0, \dots, n-1$. Edge $\eta_i$ is represented by vector
\begin{equation}
\bm{\eta}_i = \bm{\nu}_{i+1} - \bm{\nu}_i,
\end{equation}
in which $\bm{\nu}_i$ is the position vector of $\nu_{i}$. The discretized arc length $r_i$ at $\nu_i$ is computed with
\begin{equation}
r_i=\begin{dcases}
    0&  \text{if $i = 0$} \quad \text{and}\\
    \sum_{j = 0}^{i-1} | \bm{\eta}_{j} |&  \text{if $i >0$},
\end{dcases}
\label{eq:acr_len_sim}
\end{equation}
 where $r_{n-1}$ represents $r_\omega$. The unit normal $\bm{e}_i$ of $\eta_i$ that points outward from $\Omega_\alpha$ is computed as
\begin{equation}
\bm{e}_i = \mathcal{R} \frac{\bm{\eta}_i}{| \bm{\eta}_i |}, \quad \text{with} \quad 
\mathcal{R} = 
\begin{bmatrix}
0 & -1\\
1 & 0
\end{bmatrix},
\end{equation}
where $\mathcal{R}$ rotates any vector $90^{\circ}$ counterclockwise in the $yz$ plane. Following \citet{Gouraud1971}, the outward surface unit normal $\bm{n}_i$ at node $\nu_i$ is calculated as
\begin{equation}
\bm{n}_i = \begin{dcases}
    \bm{k}&  \text{if $i = 0$},\\
    \frac{\bm{e}_{i-1} + \bm{e}_i}{| \bm{e}_{i-1} + \bm{e}_i |}&  \text{if $0< i < n-1$}, \quad \text{and}\\
    \bm{\jmath}& \text{if $i = n-1$}.
\end{dcases}
\label{eq:norm_sim}
\end{equation}
By \eqref{eq:gr_rts}, growth at $\nu_i$ is expressed as
\begin{equation}
\Delta g_i = \frac{\sigma_i}{\sigma_\alpha}\Delta g_\alpha,
\label{eq:d_sim}
\end{equation}
and the position vector $\hat{\bm{\nu}}_i$ of $\nu_i$ after a growth step can be calculated as
\begin{equation}
\hat{\bm{\nu}}_i = \Delta g_i\bm{n}_i + \bm{\nu}_i.
\label{eq:node_disp}
\end{equation}

To maintain a well-resolved profile during growth, nodes are added when edge lengths exceed $| \bm{\eta}_\text{max} |$ and removed when they fall below $| \bm{\eta}_\text{min} | = 0.49 | \bm{\eta}_\text{max} |$. Growth step $\Delta g_\alpha$ is calculated as  
\begin{equation}
\Delta g_\alpha = \frac{\sigma_\alpha}{\sigma_\text{max}} | \bm{\eta}_\text{min} | ,
\label{eq:delta_g}
\end{equation}
where $\sigma_\text{max}$ denotes the largest value of $\sigma_i$. For the last growth step, $\Delta g_\alpha$ is set so that $g_\alpha = g_e$. During growth, nodes are prevented from crossing the $y$ and the $z$ axes, and topological merging is allowed, meaning that two points on $\partial\Omega_\alpha$ can meet without intersecting. The simulation algorithms and implementations are delineated in Appendices B and C, respectively.

\section{Fitting approach and procedure}
\label{sec:fit}
In our fitting approach, we assume that $r_\omega/d_\alpha$ is sufficiently large, enabling us to use $d_\alpha$ and $\acute{\sigma}_\omega/\sigma_\alpha$ as the parameters of our model. Furthermore, we suppose that the profiles of $\partial \Omega_\alpha (g_\alpha = 0)$ and $\partial \Omega_\alpha (g_\alpha = g_e)$ are given. As delineated in \secref{sec:theoretical_model}, $g_e$ can be obtained from the given profiles. With the profile of $\partial \Omega_\alpha (g_\alpha = 0)$ and initial guesses for $d_\alpha$ and $\acute{\sigma}_\omega/\sigma_\alpha$, we compute \eqref{eq:node_disp} to obtain the simulated profile of $\partial \Omega_\alpha$ after one growth step. These computations are repeated until the simulated profile of $\partial \Omega_\alpha (g_\alpha = g_e)$ is obtained. Subsequently, $d_\alpha$ and $\acute{\sigma}_\omega/\sigma_\alpha$ are optimized until the simulated profile of $\partial \Omega_\alpha (g_\alpha = g_e)$ fits the given profile of $\partial \Omega_\alpha (g_\alpha = g_e)$ best. We emphasize that unique values of $d_\alpha$ cannot be determined when $\acute{\sigma}_\omega/\sigma_\alpha = 1$, as $\sigma/\sigma_\alpha$ no longer depends on $d_\alpha$, which can be derived by \eqref{eq:res_sol_r_D}. The numerical implementation of our approach is delineated in Appendix C.

In the fitting procedure, a measure for the distance between two profiles is minimized. We achieve this by remeshing each profile with $n_r$ nodes. These nodes are placed on the original profiles, and the distance between consecutive nodes, as measured along the original profiles, is fixed. The sum of the squared distances between nodes with the same index is minimized.

\section{Demonstration}
\label{sec:demo}
\subsection{Setup, outline, and motivation}
\begin{figure*}
    \centering
    \includegraphics{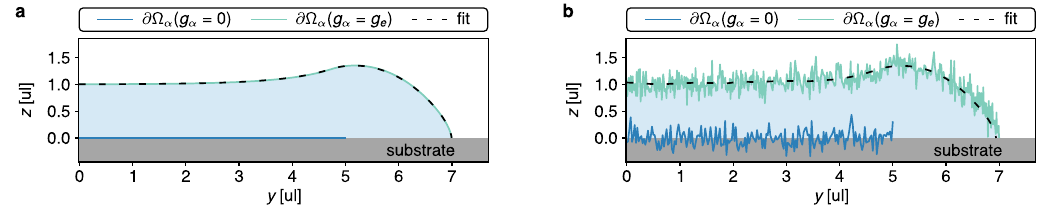}
    \caption{{\bf Demonstration.} (a) The blue and green curves represent given profiles used to demonstrate our fitting approach. The green curve is simulated with parameters $d_\alpha = 1$ unit of length (ul), $\acute{\sigma}_\omega/\sigma_\alpha = 2$, $g_e = 1~\mathrm{ul}$, and $n(g_\alpha=0) = 1600$. The dashed curve indicates the best-fitting profile obtained by the approach. The effects of profile resolution and remeshing on the relative errors of the fitting parameters are shown in Fig.~\ref{fig:resol_remesh}a,b, while CPU time is detailed in Fig.~\ref{fig:resol_remesh}c. (b) The blue and green curves from (a) are revisited with added noise. In this case, noise is obtained from a Gaussian distribution with a standard deviation SD equal to 0.15~ul. The smoothed versions of the curves serve as given profiles for the fitting approach. As before, the dashed curve shows the best-fitting profile. The influence of noise and smoothing on the relative errors of the fitting parameters is presented in Fig.~\ref{fig:noise_smooth}.}
\label{fig:vis_demo}
\end{figure*}
\begin{figure*}
    \centering
    \includegraphics{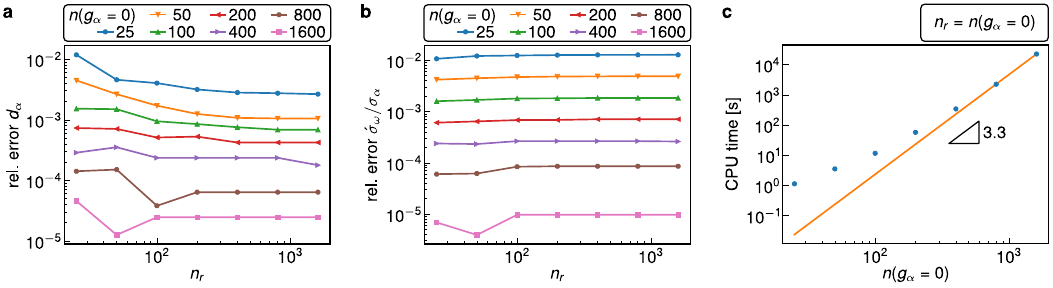}
    \caption{{\bf Resolution and remeshing.} (a, b) Relative errors in the fitting parameters $d_\alpha$ and $\acute{\sigma}_\omega/\sigma_\alpha$ as a function of $n_r$, the number of nodes used to fit the simulated profiles to the given profiles, for the growth delineated in Fig.~\ref{fig:vis_demo}a. The simulated profiles are produced for various values of $n(g_\alpha = 0)$. (c) CPU time required to fit simulated profiles to given profiles as a function of $n(g_\alpha = 0)$, with $n_r = n(g_\alpha = 0)$. The straight line represents an exponential fit to the data, indicating a power-law relationship with an exponent of approximately 3.3.}
   \label{fig:resol_remesh}
\end{figure*}
To demonstrate our fitting approach, we focus on growth initiating from a profile resembling a flat strip with $r_\omega = 5$ units of length (ul). Growth occurs under conditions $d_\alpha = 1$~ul and $\acute{\sigma}_\omega/\sigma_\alpha = 2$ and terminates when $g_e = 1$~ul. With these conditions, we produce the profile of $\partial \Omega_\alpha (g_\alpha = 0)$ and simulate the profile of $\partial \Omega_\alpha (g_\alpha = g_e)$ with $n (g_\alpha = 0) = 1600$. When using our fitting approach, these profiles serve either directly as the given profiles or as the basis for generating lower-resolution and noisy versions. The values of $| \bm{\eta}_\text{max} |$ are consistently calculated as $5~\text{ul}/ n(g_\alpha = 0)$, and during fitting, the measure for the distance between profiles is minimized using the Nelder--Mead method \cite{Gao2012}. The initial guesses for fitting are fixed at $d_\alpha = 0.5$~ul and $\acute{\sigma}_\omega/\sigma_\alpha = 2.5$, and computations are performed on a single CPU.

We plot the given profiles generated with $n(g_\alpha = 0) = 1600$ in Fig.~\ref{fig:vis_demo}(a). The dashed curve represents the simulated profile of $\partial \Omega_\alpha (g_\alpha = g_e)$ that fits the given profile of $\partial \Omega_\alpha (g_\alpha = g_e)$ best as obtained using our approach. The impact of boundary resolution and remeshing on the relative errors of the fitting parameters and the CPU time are discussed in \secref{subsec:resol_rem}. Fig.~\ref{fig:vis_demo}(b) illustrates the application of our approach to fit a simulated profile to a noisy given profile. The effects of noise and smoothing on the fitting process are analyzed in \secref{subsec:noise}.

Based on experimental results, we conclude that there is evidence for the bump-like feature shown in Fig.~\ref{fig:vis_demo}, which motivates this demonstration. \citet{Sartori2019} observed similar features in a diamond disk grown via chemical vapor deposition on a silicon substrate using an ultra-thin nanodiamond seed layer. Our simulations reproduce this feature only when the substrate acts as a source, as demonstrated by \citet{Cortes2024}, who provided evidence that a silicon substrate can behave this way.

\subsection{Resolution and remeshing}
\label{subsec:resol_rem} 
We evaluate the effect of $n(g_\alpha = 0)$, which represents the resolution of our simulation, and $n_r$, the number of nodes used for remeshing, on the accuracy of $d_\alpha$ and $\acute{\sigma}_\omega/\sigma_\alpha$ as found by our fitting approach. The relative errors on $d_\alpha$ and $\acute{\sigma}_\omega/\sigma_\alpha$ obtained are plotted in Fig.~\ref{fig:resol_remesh}a,b. For $n(g_\alpha = 0) = 1600$, the errors are on the order of $10^{-5}$ and are mainly caused by the remeshing process. For $n(g_\alpha = 0) = 200$ and below, the errors are mainly caused by lowering the resolution, and for $n_r = 200$ and above, the errors tend to stabilize. In practical scenarios, we assume that significantly increasing $n_r$ beyond the value $n(g_\alpha = 0)$ of a given profile will not result in a reduction of the error on $d_\alpha$, provided that $n(g_\alpha = 0)$ is of the same order of magnitude as $n(g_\alpha = g_e)$.

To investigate CPU time, we evaluate the time it takes to fit a simulated profile to a given profile. Both types of profiles are produced with $n(g_\alpha = 0) = n_r$. As shown in Fig.~\ref{fig:resol_remesh}c, CPU time tends to scale with a power of approximately 3.3 as a function of $n(g_\alpha = 0)$. For $n(g_\alpha = 0) = 200$, one fitting procedure can be carried out within a minute. In this situation, the relative errors on the fitting parameters remain below $10^{-3}$, which we expect is low compared to experimental errors.

\subsection{Noise and smoothing}
\label{subsec:noise}
\begin{figure*}
    \centering
    \includegraphics{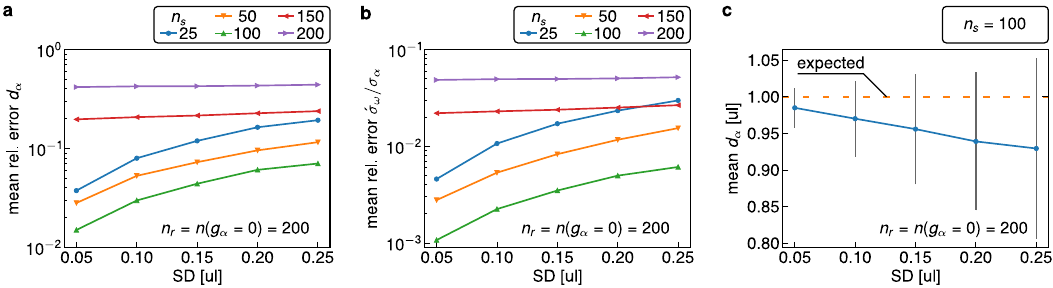}
    \caption{{\bf Noise and smoothing.} (a, b) Mean relative errors in the fitting parameters $d_\alpha$ and $\acute{\sigma}_\omega/\sigma_\alpha$ as a function of SD, the standard deviation representing noise, for the growth delineated in Fig.~\ref{fig:vis_demo}b. Remeshing is performed with $n_r = 200$, and the simulated profiles are produced for $n(g_\alpha = 0) = 200$ and various values of $n_s$, the window of the Savitzky--Golay filter \cite{Savitzky1964} used for smoothing. (c) The values of $d_\alpha$ as a function of SD with $n_s = 100$. The error bars represent the standard deviations in $d_\alpha$, and the dashed line is the expected value of $d_\alpha$.}
   \label{fig:noise_smooth}
\end{figure*}
To assess the robustness of our fitting approach for use with experimentally obtained given profiles, we investigate the effect of profile noise and its smoothing on the accuracy of fitting parameters $d_\alpha$ and $\acute{\sigma}_\omega/\sigma_\alpha$. Guided by our findings in \secref{subsec:resol_rem}, simulated profiles are produced with $n (g_\alpha = 0) = 200$, and remeshing is done with $n_r = 200$. After remeshing the profiles of $\partial \Omega_\alpha (g_\alpha = 0)$ and $\partial \Omega_\alpha (g_\alpha = g_e)$ generated with $n(g_\alpha = 0) = 1600$, we add noise to those profiles by modifying the $z$ coordinates of nodes. We do this using randomly obtained values from Gaussian distributions centered at zero with standard deviation (SD) values ranging from 0.05~ul to 0.25~ul. The profiles are smoothed by a quadratic Savitzky–Golay filter \cite{Savitzky1964} with window $n_s$ values ranging from 25 to 200. After smoothing, all negative $z$-values of the profiles are set to zero. These profiles serve as given profiles for our fitting approach.

Fig.~\ref{fig:noise_smooth}a,b contains the mean relative errors of $d_\alpha$ and $\acute{\sigma}_\omega/\sigma_\alpha$ versus SD. These values are obtained from $10^3$ pairs of given profiles. In all cases, we observe that the errors increase as SD increases. As $n_s$ increases, the errors first decrease, and for the $n_s$ values larger than 100 increase. The reason for this increase is that the given profiles are smoothed excessively. By transforming $n_s$ into length $l_s = r_\omega n_s / n$, we find that at $g_\alpha = 0$, $l_s (n_s = 100) = 2.5$~ul. From this, we infer that smoothing is excessive if $l_s$ is similar to or larger than the bump-like feature observed in the profiles depicted in Fig.~\ref{fig:vis_demo}a. From a practical perspective, this means that $n_s$ should be less than the number of nodes contained in the bump-like feature. Fig.~\ref{fig:noise_smooth}c contains the mean values of $d_\alpha$ with error bars denoting the standard deviations of $d_\alpha$. As SD increases, we observe a greater underestimation of the mean values. This effect may be caused by noise shifting the profiles in the positive direction of the $z$ axis. The mean CPU time required to complete each calculation is approximately 45~s.

\section{Conclusions}
To gain insight into the complex surface processes involved in thin-film synthesis within a reactive fluid, we devised a fitting approach to probe the diffusion length of growth precursors, $d_\alpha$. This was achieved by leveraging intersurface diffusion between a film strip and the underlying substrate. Our approach integrates the development of a theoretical growth model, a simulation strategy, and a fitting procedure.

Since the substrate serves as a source or sink for growth precursors, the growth of the strip becomes inhomogeneous, resulting in a distinct strip profile. Within the framework of our theoretical model, we demonstrated that for strips significantly wider than $d_\alpha$, only two parameters influence the growth of the strip: $\acute{\sigma}_\omega / \sigma_\alpha$ and $d_\alpha$. Here, $\acute{\sigma}_\omega / \sigma_\alpha$ represents the scaled concentration of adsorbed precursors at the strip–substrate–fluid triple line relative to the center of the strip.

In our growth simulation strategy, the strip profile was modeled as a chain of nodes. Growth was represented by the displacement of these nodes while maintaining an approximately constant node density. The displacement was proportional to the concentration $\sigma$ of adsorbed growth precursors at each node, determined using our theoretical model.

Using our fitting approach and procedure, we showed that starting from a pre-growth profile, a simulated profile can be fitted to a post-growth profile. The fitting parameters $\acute{\sigma}_\omega / \sigma_\alpha$ and $d_\alpha$ are optimized to minimize discrepancies between the simulated and given profiles.

We examined the scenario wherein the substrate supplies growth precursors, creating a bump-like feature at the edge of a strip. The influence of chain resolution on the fitting parameters and CPU time was analyzed. From our results, we infer that relative errors in the fitting parameters remain below $10^{-3}$ for computations completed within one minute on a single CPU.

Given that experimental profiles often contain noise, we tested the robustness of our fitting approach using artificially generated noisy profiles. We analyzed how the noise level and noise smoothing affected errors in the fitting parameters. Using a Savitzky--Golay filter \cite{Savitzky1964} for noise smoothing, we found that the smoothing window should be smaller than the number of nodes comprising the bump-like feature at the strip edge.

Our fitting approach is a novel tool to bridge the gap between experiment and theory. Film strips can be fabricated by processing thin films using photolithography and reactive ion etching, and profiles pre- and post-growth can be obtained by atomic force microscopy. With our approach, the effect of growth parameters on surface diffusion lengths can then be investigated.

\subsection*{CRediT authorship contribution statement}
{\bf Stoffel D.\ Janssens:} conceptualization, formal analysis, data curation, investigation, methodology, software, writing -- original draft. {\bf Francisco S.\ Forte Neto:} conceptualization, methodology, writing -- review \& editing. {\bf David V\'azquez-Cort\'es:} conceptualization, writing -- review \& editing. {\bf Fernando P.\ Duda:} formal analysis, methodology, writing -- review \& editing. {\bf Eliot Fried:} conceptualization, formal analysis, methodology, project administration, resources, writing -- review \& editing.

\subsection*{Declaration of competing interest}
The authors declare that they have no known competing financial interests or personal relationships that could have appeared to influence the work reported in this paper.

\section*{Data availability}
All data, code to generate the data, code to plot the data, and numerical implementation are available as Supplementary material.

\subsection*{Acknowledgments}
This project is supported by OIST, with subsidy funding from the Cabinet Office, Government of Japan. The project is also supported by Coordena\c{c}\~{a}o de Aperfei\c{c}oamento de Pessoal de N\'{i}vel Superior -- Brasil (CAPES) -- Finance Code 001, through a CAPES-PrInt scholarship awarded to Francisco S.\ Forte Neto for conducting part of his doctoral studies at OIST.

\makeatletter
\renewcommand{\@seccntformat}[1]{}
\makeatother

\appendix
\section*{Appendix A. Theory}
\label{sec:theory}
\setcounter{equation}{0}
\renewcommand{\theequation}{A.\arabic{equation}}
The reaction rates on $\partial\Omega_\alpha$ are specified as
\begin{equation}
J_1^+ = k_1^+ \sigma_\gamma \sigma^\text{V}_\alpha, \quad 
J_1^- = k_1^- \sigma, \quad \text{and} \quad
J_2^+ = k_2^+ \sigma,
\label{eq:fluxes_D}
\end{equation}
in which $\sigma_\gamma$ and $\sigma^\text{V}_\alpha$ are the concentrations of P$_\gamma$ and V$_\alpha$, respectively, and $k_1^+$, $k_1^-$, and $k_2^+$ are rate constants. The reaction rates on $\partial\Omega_\beta$ are specified as
\begin{equation}
J_3^+ = k_3^+ \sigma_\gamma \sigma^\text{V}_\beta \quad \text{and} \quad
J_3^- = k_3^- \sigma,
\label{eq:fluxes_S}
\end{equation}
respectively, in which $\sigma^\text{V}_\beta$ is the concentration of V$_\beta$ and $k_3^+$ and $k_3^-$ are rate constants. With constants $\sigma^\text{S}_\alpha$ and $\sigma^\text{S}_\beta$ denoting the sum of the occupied and unoccupied vacancy concentrations on $\partial \Omega_\alpha$ and $\partial \Omega_\beta$, respectively, we find
\begin{align}
\sigma^\text{S}_\alpha &= \sigma^\text{V}_\alpha + \sigma \quad \text{and}
\label{eq:SD}\\
\sigma^\text{S}_\beta &= \sigma^\text{V}_\beta + \sigma.
\label{eq:SS}
\end{align}
By \eqref{eq:KPP_D}, \eqref{eq:KPP_S}, \eqref{eq:fluxes_D}, \eqref{eq:fluxes_S}, \eqref{eq:SD}, and \eqref{eq:SS}, we obtain 
\begin{align}
D_\alpha \frac{\partial^2 \sigma}{\partial r^2} - a_\alpha\sigma + b_\alpha &= 0
&\text{on} \quad &\partial \Omega_\alpha \quad &\text{and}
\label{eq:KPP_Dfin}\\
D_\beta \frac{\partial^2 \sigma}{\partial r^2} - a_\beta\sigma + b_\beta &= 0
&\text{on} \quad &\partial \Omega_\beta.
\label{eq:KPP_Sfin}
\end{align}
in which
\begin{align}
a_\alpha &= k_1^+ \sigma_\gamma + k_1^- + k_2^+,
& a_\beta &= k_3^+ \sigma_\gamma + k_3^-, \label{a_alpha_beta}\\
b_\alpha &= k_1^+ \sigma_\gamma \sigma^\text{S}_\alpha \quad \text{and},
&b_\beta &= k_3^+ \sigma_\gamma\sigma^\text{S}_\beta.
\end{align}
Relations \eqref{eq:KPP_Dfin} and \eqref{eq:KPP_Sfin} are second-order linear ordinary differential equations that we solve by following the method of combination of variables \cite{Boltzmann1894}. We combine the variables as
\begin{align}
f_\alpha &= \sigma - \sigma_\alpha,
\quad \sigma_\alpha = \frac{b_\alpha}{a_\alpha}, 
\quad d_\alpha = \sqrt{\frac{D_\alpha}{a_\alpha}},
\label{eq:subsD}\\
f_\beta &= \sigma - \sigma_\beta, 
\quad \sigma_\beta = \frac{b_\beta}{a_\beta}, \quad \text{and} \quad
\quad d_\beta = \sqrt{\frac{D_\beta}{a_\beta}}.
\label{eq:subsS}
\end{align}
With \eqref{eq:KPP_Dfin} and \eqref{eq:subsD} and relations \eqref{eq:KPP_Sfin} and \eqref{eq:subsS}, we obtain
\begin{align}
\frac{\partial^2 f_\alpha}{\partial r^2} - \frac{1}{d_\alpha^2} f_\alpha &=0 \quad \text{and} \label{eq:MBDED}\\
\frac{\partial^2 f_\beta}{\partial r^2} - \frac{1}{d_\beta^2} f_\beta &= 0, \label{eq:MBDES}
\end{align}
reaspectively. The general solutions of \eqref{eq:MBDED} and \eqref{eq:MBDES} are
\begin{align}
f_\alpha &= A_\alpha \exp{\left(\frac{r}{d_\alpha}\right)} + B_\alpha \exp{\left(-\frac{r}{d_\alpha}\right)} \quad \text{and} \label{eq:GMBFD}\\
f_\beta &= A_\beta \exp{\left(\frac{r}{d_\beta}\right)} + B_\beta \exp{\left(-\frac{r}{d_\beta}\right)} \label{eq:GMBFS},
\end{align}
in which $A_\alpha$, $B_\alpha$, $A_\beta$, and $B_\beta$ are constants. The conditions that we apply to obtain these constants are
\begin{align}
\lim_{r \downarrow 0} \frac{\partial \sigma }{\partial r} &= 0, \quad 
&\lim_{r \uparrow r_\omega} \sigma &= \sigma_\omega,
\label{eq:BC12}\\
\lim_{r \downarrow r_\omega} \sigma &= C\sigma_\omega, \quad \text{and} \quad
&\lim_{r \uparrow \infty}\sigma &= \sigma_\beta.
\label{eq:BC34}
\end{align}
Relation~\eqref{eq:BC12}$_1$ is a consequence of the symmetry of the strip, and \eqref{eq:BC12}$_2$ describes the precursor concentration on $\partial\Omega_\alpha$ at $\omega$. For obtaining \eqref{eq:BC34}$_1$, we assume that the chemical potentials of the precursors on $\partial\Omega_\alpha$ and $\partial\Omega_\beta$ are
\begin{align}
\mu_\alpha &= \mu^0_\alpha + k_BT\ln{\sigma} \quad \text{and} \label{eq:chem_pot_a}\\
\mu_\beta &= \mu^0_\beta + k_BT\ln{\sigma},  \label{eq:chem_pot_b}
\end{align}
respectively, in which $T$ denotes the absolute temperature and $k_B$ is the Boltzmann constant \cite{Saggion2020}. By \eqref{eq:BC12}$_2$, \eqref{eq:chem_pot_a}, \eqref{eq:chem_pot_b}, and taking $\mu_\alpha = \mu_\beta$ at $r_\omega$, \eqref{eq:BC34}$_1$ is obtained, in which
\begin{equation}
C = \exp{\left(\frac{\mu^0_\alpha-\mu^0_\beta}{k_BT}\right)}.
\end{equation}
By \eqref{eq:KPP_Sfin}, \eqref{eq:subsS}$_2$, \eqref{eq:lim_con_beta}, and realizing that
\begin{equation}
\lim_{r \uparrow\infty} \frac{\partial \sigma}{\partial r} = 0,
\label{eq:lim_con_beta}
\end{equation}
we find \eqref{eq:BC34}$_2$. Relation \eqref{eq:BC34}$_2$ shows that $\sigma_\beta$ is the precursor concentration on $\partial\Omega_\beta$ far away from $\omega$. To find $A_\alpha$ and $B_\alpha$, we transform \eqref{eq:GMBFD} with \eqref{eq:subsD} into
\begin{equation}
\sigma = \sigma_\alpha + A_\alpha \exp{\left(\frac{r}{d_\alpha}\right)} + B_\alpha \exp{\left(-\frac{r}{d_\alpha}\right)} \label{eq:sol_r_D_init},
\end{equation}
and by solving \eqref{eq:BC12}$_1$ with \eqref{eq:sol_r_D_init}, it follows that $A_\alpha = B_\alpha$. With \eqref{eq:BC12}$_2$ and \eqref{eq:sol_r_D_init}, we obtain
\begin{equation}
\begin{aligned}
\sigma =&~ \sigma_\alpha + \frac{\sigma_\omega - \sigma_\alpha}{\exp{\left(\dfrac{2r_\omega}{d_\alpha}\right)}+1}\\
&~\left[\exp{\left(\frac{r_\omega + r}{d_\alpha}\right)} + \exp{\left(\frac{r_\omega - r}{d_\alpha}\right)} \right] 
\label{eq:sol_r_D},
\end{aligned}
\end{equation}
on $\Omega_\alpha$, which is equivalent to
\begin{equation}
\sigma = \sigma_\alpha + \frac{2(\sigma_\omega - \sigma_\alpha)\exp{\left(\dfrac{r_\omega}{d_\alpha}\right)}}
{\exp\left(\dfrac{2r_\omega}{d_\alpha}\right)+1}
\cosh{\left(\frac{r}{d_\alpha}\right)}
\label{eq:sol_r_D_cosh}.
\end{equation}
To find $A_\beta$ and $B_\beta$, we transform \eqref{eq:GMBFS} with \eqref{eq:subsS} into
\begin{equation}
\sigma = \sigma_\beta + A_\beta \exp{\left(\frac{r}{d_\beta}\right)} + B_\beta \exp{\left(-\frac{r}{d_\beta}\right)}. 
\label{eq:sol_r_S_init}
\end{equation}
With \eqref{eq:BC34}$_2$ and \eqref{eq:sol_r_S_init}, we find that $A_\beta = 0$. With \eqref{eq:BC34}$_1$ and \eqref{eq:sol_r_S_init}, we obtain
\begin{equation}
\sigma = \sigma_\beta + \left(C\sigma_\omega - \sigma_\beta\right)
\exp{\left(\frac{r_\omega - r}{d_\beta}\right)},
\label{eq:sol_r_S}
\end{equation}
on $\Omega_\beta$. The condition to couple $\Omega_\alpha$ and $\Omega_\beta$ assures that mass accumulation is prevented at $\omega$. Under the quasi-steady-state assumption, this condition is expressed as
\begin{equation}
\lim_{r \downarrow r_\omega}D_\beta \frac{\partial \sigma }{\partial r} = \lim_{r \uparrow r_\omega}D_\alpha \frac{\partial \sigma }{\partial r},
\label{eq:BC3}
\end{equation}
which means that the flux of adsorbed precursors is the same on each side of $\omega$. By solving \eqref{eq:BC3} with \eqref{eq:sol_r_D} and \eqref{eq:sol_r_S}, we find that
\begin{equation}
\sigma_\omega = \frac{\sigma_\beta+\sigma_\alpha\dfrac{d_\beta}{d_\alpha}\dfrac{D_\alpha}{D_\beta}\tanh{\left(\dfrac{r_\omega}{d_\alpha}\right)}} 
{C+\dfrac{d_\beta}{d_\alpha}\dfrac{D_\alpha}{D_\beta}\tanh{\left(\dfrac{r_\omega}{d_\alpha}\right)}}. 
\label{eq:bcon}
\end{equation}

By defining diffusion lengths as lengths that naturally appear by solving \eqref{eq:MBDED} and \eqref{eq:MBDES}, we infer from \eqref{eq:sol_r_D} and \eqref{eq:sol_r_S} that $d_\alpha$ and $d_\beta$ are the diffusion lengths of adsorbed precursors on $\partial\Omega_\alpha$ and $\partial\Omega_\beta$, respectively. Since $d_\alpha$ and $d_\beta$ can be expressed as 
\begin{equation}
d_\alpha = \sqrt{D_\alpha \tau_\alpha} \quad \text{and} \quad d_\beta = \sqrt{D_\beta \tau_\beta},
\label{eq:diff_len}
\end{equation}
where $\tau_\alpha$ and $\tau_\beta$ are the lifetimes of precursors adsorbed to $\partial\Omega_\alpha$ and $\partial\Omega_\beta$, respectively, we find with \eqref{eq:subsD}$_3$ and \eqref{eq:subsS}$_3$ that 
\begin{equation}
\tau_\alpha = \frac{1}{a_\alpha} \quad \text{and} \quad \tau_\beta = \frac{1}{a_\beta}.
\label{eq:eff_lifetimes}
\end{equation}

By evaluating \eqref{eq:sol_r_D_cosh} and \eqref{eq:bcon} with conditions
\begin{equation}
\lim_{r_\omega/d_\alpha \uparrow \infty} \sigma = \acute{\sigma} \quad \text{and} \quad
\lim_{r_\omega/d_\alpha \uparrow \infty} \sigma_\omega = \acute{\sigma}_\omega,
\end{equation}
we obtain
\begin{align}
\acute{\sigma}(r = 0) & = \sigma_\alpha
\label{eq:sol_r_D_as} \quad \text{and}\\
\acute{\sigma}_\omega &= \frac{\sigma_\beta+\sigma_\alpha\dfrac{d_\beta}{d_\alpha}\dfrac{D_\alpha}{D_\beta}} 
{C+\dfrac{d_\beta}{d_\alpha}\dfrac{D_\alpha}{D_\beta}},
\label{eq:bcon_as}
\end{align}
respectively. By evaluating \eqref{eq:sol_r_D} and \eqref{eq:bcon} with conditions
\begin{equation}
\lim_{r_\omega/d_\alpha \downarrow 0} \sigma = \grave{\sigma} \quad \text{and} \quad
\lim_{r_\omega/d_\alpha \downarrow 0} \sigma_\omega = \grave{\sigma}_\omega,
\end{equation}
we find that 
\begin{equation}
\grave{\sigma} = \grave{\sigma}_\omega = \frac{\sigma_\beta}{C},
\label{eq:sigma_grave}
\end{equation}
which means that the concentration of adsorbed precursors on a relatively thin strip is approximately $\sigma_\beta/C$.

We use a scaling strategy in which concentration, length, and time are divided by $\sigma_\alpha$, $d_\alpha$, and $\tau_\alpha$, respectively. Scaled quantities are marked with $^*$ so that the scaled counterparts of \eqref{eq:sol_r_D_cosh}, \eqref{eq:sol_r_S} and \eqref{eq:bcon} are
\begin{align}
\sigma^* & = 1 + \frac{2(\sigma^*_\omega - 1)\exp{(r^*_\omega)}}{\exp{(2r^*_\omega)}+1}
\cosh{(r^*)},
\label{eq:sol_r_D_cosh_sc}\\
\sigma^* & = \sigma^*_\beta + (C\sigma^*_\omega - \sigma^*_\beta)\exp{\left(\frac{r^*_\omega - r^*}{d^*_\beta}\right)}, \quad \text{and}
\label{eq:sol_r_S_sc}\\
\sigma^*_\omega & = \frac{\sigma^*_\beta + d^*_\beta D\tanh{(r^*_\omega)}}
{C + d^*_\beta D\tanh{(r^*_\omega)}},
\label{eq:bcon_sc}
\end{align}
respectively, with $D = D_\alpha / D_\beta$.
If $\sigma^*_\omega > 1$, the substrate is a source for growth precursors, and if $\sigma^*_\omega < 1$, it becomes a sink. The boundary layer length is naturally defined as the sum of $d^*_\alpha$ and $d^*_\beta$, in which $d^*_\alpha = 1$. The scaled counterparts of asymptotic relations \eqref{eq:sol_r_D_as}, \eqref{eq:bcon_as}, and \eqref{eq:sigma_grave} become
\begin{align}
\acute{\sigma}^*(r^*=0) & = 1,
\label{eq:sol_r_D_acute_sc}\\
\acute{\sigma}^*_\omega & = \frac{\sigma^*_\beta + d^*_\beta D}{C + d^*_\beta D}, \quad \text{and}
\label{eq:bcon_acute_sc}\\
\grave{\sigma}^* & = \frac{\sigma^*_\beta}{C},
\label{eq:sigma_grave_sc}
\end{align}
respectively.

\section*{Appendix B. Algorithms}
\setcounter{equation}{0}
\renewcommand{\theequation}{B.\arabic{equation}}
\label{sec:algorithms}

After simulating a growth step, the following algorithms are executed sequentially to prevent nodes from crossing the $x$ and $y$ axes, to prevent edge lengths from becoming large or small, and to account for topological merging.
\begin{enumerate}[(a)]
\item \emph{Crossing the $z$ axis.} If nodes with negative $y$ values exist, the straight line equation defined by the highest index node with a negative $y$ value and the subsequent node, which has a positive $y$ value, is computed. All nodes with negative $y$ values are then replaced with one node that is placed at the intersection of the computed line and the $z$ axis.
\item \emph{Crossing the $y$ axis.} If nodes with negative $z$ values exist, the straight line equation defined by the lowest index node with a negative $z$ value and the preceding node, which has a positive $z$ value, is computed. All nodes with negative $z$ values are then replaced with one node that is placed at the intersection of the computed line and the $x$ axis.
\item \emph{Topological merging.} The distances between all nodes are calculated, and the node pairs with a distance below a predetermined lower bound $| \bm{\eta}_\text{min} |$ are stored. Stored node pairs that coincide with edges are filtered, and the first remaining node pair is selected. Then, all the nodes with an index between the indices of that selected pair are removed. The selected pair is replaced by one node located in the middle of that pair. This procedure is repeated until all internode distances that do not coincide with edge lengths are above $| \bm{\eta}_\text{min} |$. The removed nodes can be saved to specify the location of a cavity. This algorithm is computationally demanding, which is typical for a numerical implementation that deals with topological changes \cite{Vetter2024}.
\item \emph{Lower-bound edge length.} Nodes are deleted to form edges with a length above $| \bm{\eta}_\text{min} |$. Nodes $\nu_1$ and $\nu_n$ are not removed. Details regarding this algorithm are provided in the code of the numerical implementation.
\item \emph{Upper-bound edge length.} Edge lengths are computed, and edges with a length over a predetermined upper bound $| \bm{\eta}_\text{max} |$ are selected. A node is added in the middle of each selected edge.\\
\end{enumerate}

\section*{Appendix C. Numerical implementation}
\setcounter{equation}{0}
\renewcommand{\theequation}{C.\arabic{equation}}
\setcounter{figure}{0}
\renewcommand{\thefigure}{C.\arabic{figure}}
\label{sec:software}
\begin{figure}
    \centering
    \includegraphics{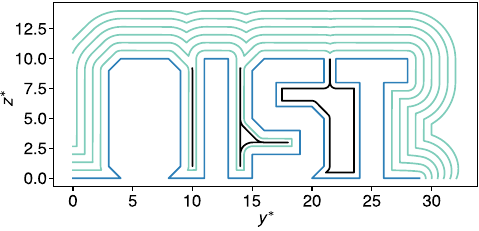}
    \caption{{\bf Numerical implementation.} Graph similar to the one in the PDF file generated by our numerical implementation of the growth simulation strategy. The dark blue curve represents $\partial\Omega^*_\alpha(g^*_\alpha= 0)$, the light green curves represent $\partial\Omega_\alpha^*$ at equally-spaced values of $g^*$, including value $g^*_e$. The black curves denote portions of $\partial\Omega^*_\alpha$ that are deleted by the \emph{Topological merging} algorithm.}
\label{fig:example_soft}
\end{figure}
A numerical implementation of the growth simulation strategy delineated in \secref{sec:sim_strat} is provided as Supplementary material. The implementation is written in Python and relies on the Numpy \cite{Harris2020}, Scipy \cite{Virtanen2020}, Matplotlib, and Math packages. Scaled relations are used, and $\sigma^*_\beta$, $d^*_\beta$, $D$, $C$, $g^*_e$, and $| \bm{\eta}^*_\text{max} |$ should be set in the code. A Numpy array of data type float with nodes representing the profile of $\partial\Omega^*_\alpha(g^*_\alpha= 0)$ should also be set. The implementation automatically alters the number of nodes using input $| \bm{\eta}^*_\text{max} |$ to set the resolution of the simulation. The node array representing the profile of $\partial\Omega^*_\alpha(g^*_\alpha= g^*_e)$ and a data plot are saved as a Numpy file and a PDF file, respectively. The plot generated by the implementation is similar to Fig.~\ref{fig:example_soft} if none of the arbitrarily chosen settings are changed. In Fig.~\ref{fig:example_soft}, the dark blue curve represents $\partial\Omega^*_\alpha(g^*_\alpha= 0)$, and the light green curves represent $\partial\Omega^*_\alpha$ at equally-spaced values of $g^*$, including $g^*_\alpha= g^*_e$. The number of light green curves can be set. The black line represents portions of the profile of $\partial\Omega^*_\alpha$ that the \emph{Topological merging} algorithm removes. A black line is plotted if the number of removed nodes is larger than a value that can be set. We have included this functionality because, in some cases, the algorithm removes only a few nodes, which can be undesirable to plot. Such a case can occur at a concave corner that consists of nodes $\nu_{i-1}$, $\nu_{i}$, and $\nu_{i+1}$. Here, the distance between $\nu_{i-1}$ and $\nu_{i+1}$ is less than $| \bm{\eta}^*_\text{min} |$. Consequently, $\nu_{i}$ is removed, and $\nu_{i-1}$ and $\nu_{i+1}$ are replaced by a node that is placed in between $\nu_{i-1}$ and $\nu_{i+1}$.

A numerical implementation of the fitting approach delineated in \secref{sec:fit} is also attached to this work. The fitting implementation is written in Python and is based on a modified version of the simulation implementation. An important difference is that the fitting algorithm uses $\acute{\sigma}_\omega$ instead of $\sigma_\omega$. The fitting algorithm requires $g_e$, $| \bm{\eta}_\text{max} |$, and the unscaled profiles of $\Omega_\alpha(g_\alpha = 0)$ and $\Omega_\alpha(g_\alpha = g_e)$. Additionally, a guess for $d_\alpha$ and $\acute{\sigma}_\omega/\sigma_\alpha$ should be provided. The minimization for the fitting procedure is done with the Scipy package. The fitting algorithm provides $d_\alpha$, $\acute{\sigma}_\omega/\sigma_\alpha$, and plotted profiles saved as a PDF file.
 
For the demonstration of our fitting approach in \secref{sec:demo}, we also provide a simulation implementation for generating the given profiles. In that implementation, computations are performed with $\acute{\sigma}_\omega$ instead of $\sigma_\omega$. Adding noise and smoothing is carried out by the norm and savgol\_filter functions of the Scipy package, and CPU time is obtained with the perf\_counter() function of the time module that Python provides. Computations are performed on an Intel Xeon Gold 5218 CPU (2.30 GHz).

\section*{Supplementary material}
\href{https://github.com/StoffelJanssens/Diffusion-and-Growth}{https://github.com/StoffelJanssens/Diffusion-and-Growth}

\renewcommand{\thesection}{}


\end{document}